\documentclass[reqno]{amsart}
\usepackage{bbm}
\usepackage{amsfonts}
\usepackage{mathrsfs}
\usepackage{hyperref}
\usepackage{amssymb}  
\usepackage{CJK}

 \theoremstyle{remark}
 
 \numberwithin{equation}{section}

\begin{document}
\title[The Minimal and Maximal Sensitivity of the Simplified Weighted Sum Function]
  {The Minimal and Maximal Sensitivity of the Simplified Weighted Sum Function}

\author{Jiyou Li}
\address{Department of Mathematics, Shanghai Jiao Tong University, Shanghai, P.R. China}
\email{lijiyou@sjtu.edu.cn}

\author{Chu Luo}
\address{Center for Ubiquitous Computing, University of Oulu, Oulu, Finland} \email{chu.luo@ee.oulu.fi}

\author{Zeying Xu}
\address{Department of Mathematics, Shanghai Jiao Tong University, Shanghai, P.R. China}\email{zane\_xu@sjtu.edu.cn}


\thanks{This work is supported by the National Science Foundation of China
(11001170) and the National Science Foundation of Shanghai Municipal
(13ZR1422500).}

\begin{abstract}

Sensitivity is an important complexity measure of Boolean functions. In this paper we present properties of the minimal and maximal sensitivity of the simplified weighted sum function. A simple close formula of the minimal sensitivity of the simplified weighted sum function is obtained. A phenomenon is exhibited that the minimal sensitivity of the weighted sum function is indeed an indicator of large primes, that is, for large prime number $p$, the minimal sensitivity of the weighted sum function is always equal to one.


\end{abstract}

\maketitle \numberwithin{equation}{section}
\newtheorem{theorem}{Theorem}[section]
\newtheorem{lemma}[theorem]{Lemma}
\newtheorem{example}[theorem]{Example}
\allowdisplaybreaks

\section{Introduction}
It is well-known that Boolean functions are of most importance in the design of circuits and chips for almost all various electronic instruments. Indeed, as the digital computer system relies on the binary algebraic operations, the theory of Boolean functions are playing a more and more significant role in most areas of current and future technology, as well as in both natural science and social science, cf. (Crama and Hammer, 2011) for details. As a specific example, Boolean functions play a key role in cryptography for creating symmetric key algorithms, which is well-known closely related to number theory. The sensitivity concept of Boolean functions is originally introduced in (Cook et al., 1986). In practice, sensitivity is used as a combinatorial complexity measure of various Boolean models (Sauerhoff, 2003) (Sauerhoff and Sieling, 2005) (Canright et al., 2011) (Hatami et al., 2011).

Shparlinski (2007) showed a lower bound of the average sensitivity of the weighted sum Boolean function, also known as laced Boolean function. And he developed a conjecture about the average sensitivity. Canright et al. (2011) gave a series of formulas of the average sensitivity of the weighted sum function. Recently, Li (2012) solved the Shparlinski's conjecture by the bound on the average sensitivity of the weighted sum function. However, most existing researches focus on the average sensitivity of the weighted sum function. It is worth noting that the maximal and minimal sensitivities are also effective complexity measures of Boolean functions. This paper deals with the minimal sensitivity of the weighted sum Boolean function, which was originated from Savicky and Zak (2000) in their research of read-once branching programs and then had positions for a variety of complexity theory applications, cf. (Sauerhoff, 2003) (Sauerhoff and Sieling, 2005). Among other things, in our first main result Lemma 4.3 we obtain an amazingly simple close formula of the minimal sensitivity of the weighted sum function. In our second main result Theorem 4.4, a surprising phenomenon is found that the minimal sensitivity of the weighted sum function is indeed an indicator of large primes. That is, for prime number $p \geq 5$, the minimal sensitivity of the weighted sum function is always equal to one.

The remainder of this paper is organized as follows. In Section 2 we discuss the sensitivity of Boolean functions, especially the minimal sensitivity. Section 3 describes a new simplified weighted sum function. The main results are presented in Section 4. Finally, Section 5 concludes this paper with several open questions.

\section{Sensitivity of Boolean functions}
In this section we introduce the sensitivity of Boolean functions. For a Boolean function $f(X)$ on $n$ variables and an input
\begin{equation}
X=(x_0, x_1, \dots, x_n-1) \in \mathbb{Z}_2^n
\end{equation}
where n-dimensional space $\mathbb{Z}_m^n = \{0,1, \dots, m-1\}^n$, the sensitivity $Sen(f,X)$ denotes the number of coordinates in $X$ such that flipping one Boolean variable of $X$ will change the function value of $f(X)$. Explicitly, $Sen(f,X)$ can be given by

\begin{equation}
\sum_{i=0}^{n-1} |f(X)-f(X\oplus {e_i})|
\end{equation}
where $X\oplus {e_i}$ denotes a new vector with original Boolean variable values of $X$ and $x_i$ is flipped in the new vector.

The average sensitivity $AS(f)$ denotes the expected value of S$en(f,X)$ on every possible input $X$ over $\mathbb{Z}_2^n$. Explicitly,

\begin{equation}
AS(f)=  2^{-n} \sum_{X \in \mathbb{Z}_2^n} \sum_{i=0}^{n-1} |f(X)-f(X\oplus {e_i})|.
\end{equation}

Similarly, for every possible input $X$ over $\mathbb{Z}_2^n$, let $maxS(f)$ and $minS(f)$ be the maximal and minimal values of the sensitivity $Sen(f,X)$, respectively.

Much work (Sauerhoff and Sieling, 2005) (Shparlinski, 2007) is proposed to apply the average sensitivity of Boolean functions to practice, ranging from circuit complexity and the size of a decision tree. However, previous studies have not addressed properties of the maximal and minimal sensitivities of Boolean functions. On this basis, we focus on the minimal sensitivity of a new simplified weighted sum function.

\section{Weighted Sum Function}
Previously, the definition of the weighted sum function is proposed according to the weighted sum with a residue ring modulo a prime number. Explicitly, it can be shown in the following (Savicky and Zak, 2000).

Let $n \in \mathbb{N}^*$ and $p$ is a prime number, $p \geq n$ where no prime number $q$ meets $n \leq q < p$. For an input set $X=(x_1,x_2, \dots, x_n) \in \mathbb{Z}_2^n$, construct a function $s(X)$ by
\begin{equation}
s(X)= \sum_{k=1}^n kx_k({\rm mod}\  p), 1\leq s(X) \leq p.
\end{equation}
Then define the weighted sum function
\begin{equation}
f(X)=\left\{
\begin{array}{ll}
  x_{s(X)}, \ \  1\leq s(X)\leq n;\\
  x_1, \ \  \hbox{otherwise.}\\
    \end{array}
    \right.
\end{equation}

As the previous weighted sum function is relatively complex, a new simplified weighted sum Boolean function $f(X)$ 
can be defined as follows (Li and Luo, 2014) (Li and Luo, 2016). Let $n$
be a positive integer. For $X=(x_0,x_1,\dots,x_{n-1})\in \mathbb{Z}_2^n$, we define $s(X)$ by
\begin{equation}
s(X)= \sum_{k=0}^{n-1} kx_k({\rm mod}\  n).
\end{equation}
We then define that
\begin{equation}
f(X)=x_{s(X)}.
\end{equation}
This simplified weighted sum function is more convenient to use and compute.

\section{Our Results}
In this section we present several properties of the newly simplified weighted sum function.

\begin{theorem}
\label{theorem1}
maxS(f) = n.
\end{theorem}
\begin{proof}
Due to the specificity of the weighted sum function, this theorem is trivial. Given Boolean function $f(X)$ and its input set $X$ with $n$ Boolean variables, there always exists an input $X_1$ where $x_i = 0, i \in \{0,...,n-1\}$. In this case, $s(X_1) = 0$ and $f(X_1) = x_0=0$. For $0 \leq i<n$, $f(X_1 \oplus {e_i} ) = x_i$, which is just flipped from 0 to 1. Thus, flipping any variable in $X_1$ will change the value of $f(X_1)$. Hence, the maximum of sensitivity of the weighted sum function is $n$.
\end{proof}

\begin{theorem}
\label{theorem2}
Let $ p\geq 3 $ be any prime number. If $ p^2|n $, then $minS(f)=0$.
\end{theorem}
\begin{proof}
Assume $n=p^2q$. Consider a boolean variable input $X_1$ where $x_i=1$ if $i\equiv p-1\ ({\rm mod}\ p), 0 \leq i < n$ and $x_i=0$, otherwise. Then $s(X_1)$ can be given by
\begin{equation}
(p-1+2p-1+\cdots +p^2q-1)\ {\rm mod}\  p^2q=\frac{(p^2q^2+pq-2q)p}{2}\ {\rm mod}\  p^2q.
\end{equation}

Since $p^2q^2+pq$ is even, so is $p^2q^2+pq-2q$. So we have the conclusion that $p|s(X_1)$ and $f(X_1)=0$.

On the other hand, if $n|s(X_1)$, then we have
\[p^2q|(p^2q^2+pq-2q)p\Rightarrow pq|p^2q^2+pq-2q \Rightarrow pq|-2q\]

which is impossible. Then $p|s(X_1)$ and $s(X_1)>0$.

Assume $x_i$ is flipped.

If $ i=mp-1, m>0 $, then $s(X_1\oplus {e_i}) \neq s(X_1)$ and $ s(X_1\oplus {e_i})= s(X_1)- mp+1 + tp^2q, t=0$ or 1. Then we have $s(X_1\oplus {e_i})~{\rm mod}\ ~p=1$. Since $x_j=0, j\ {\rm mod}\ p \neq p-1, 0 \leq j < n$, $f(X_1\oplus {e_i}) = x_{s(X_1\oplus {e_i})}=0$.

If $ 0 \leq i~{\rm mod}\ p \leq p-2 $, then  $s(X_1\oplus {e_i}) = s(X_1) + i - tp^2q, t=0$ or 1. Since $0 \leq s(X_1\oplus {e_i})~{\rm mod}\ p\leq p-2$ and $x_j=0, j\ {\rm mod}\ p \neq p-1, 0 \leq j < n, j \neq i$, $f(X_1\oplus {e_i})=1$ if and only if $s(X_1\oplus {e_i}) = i$. In this case, $s(X_1)-tp^2q=0$.

Then, we have
\begin{equation}
\begin{aligned}
\frac{(p^2q^2+pq-2q)p}{2} =& mp^2q, m\in \mathbb Z\\
p^2q^2+pq-2q=&2mpq\\
p^2q+p-2=&2mp\\
p(pq+1-2m)=&2
\end{aligned}
\label{eq:jaa1}
\end{equation}
where $p$ and $(pq+1-2m)$ are two integers. Since $p\geq 3$, ~Eq.~(\ref{eq:jaa1}) is impossible. Then we have $f(X_1\oplus {e_i})=0$. Hence, $Sen(f,X_1)=0$ and $minS(f)=0$.
\end{proof}
\begin{lemma} \label{le1}

Let $k \in \mathbb{N}$ and an input $X_1$ be $(x_0,x_1,\dots,x_{n-1})$ where $x_i=1, i \in \{0, 1, ... , n-1\}$.
Then we have

 $$ Sen(f,X_1)=\left\{
\begin{aligned}
& 0, \ \  n=4k+2;\\
& 1, \ \  n=2k+1;\\
& 2,  \ \ n=4k.\\
\end{aligned}
\right.
$$
\end{lemma}
\begin{proof}
By the definition of $X_1$ we have $f(X_1) = 1$ and
\begin{equation}
s(X_1) = \frac{n(n-1)}{2} \ {\rm mod}\  n.
\end{equation}

If $n=4k+2$, $s(X_1) = 8k^2+6k+1 \ {\rm mod}\ 4k+2$. Since $8k^2+6k+1$ is odd and $4k+2$ is even, we have $s(X_1) \neq 0$. If $x_i$ is flipped, then $s(X_1\oplus {e_i})= 8k^2+6k+1+(4k+2)t-i, t \in \mathbb Z$. $f(X_1\oplus {e_i})=0$ happens if and only if $s(X_1\oplus {e_i})=i$. In this case, we have
\begin{equation}
\begin{aligned}
i=&8k^2+6k+1+(4k+2)t-i \\
2i=&8k^2+6k+1+(4k+2)t
\end{aligned}
\end{equation}
where $8k^2+6k+1$ is odd and $4k+2$ is even. There does not exist an integer $i$ in $f(X_1\oplus {e_i})=0$.
Hence, $f(X_1\oplus {e_i}) = 1$ and $Sen(f,X_1)=0=minS(f)$ if $n=4k+2$.

If $n=2k+1$, $s(X_1) = k(2k+1)\ {\rm mod}\  2k+1$. Thus we have $s(X_1)=0$. If $x_i$ is flipped, then
\begin{equation}
s(X_1\oplus {e_i})=\left\{
\begin{aligned}
& 0, \ \  i=0;\\
& n-i, \ \  \hbox{otherwise.}\\
\end{aligned}
\right.
\end{equation}
$f(X_1\oplus {e_i})=0$ happens if and only if $s(X_1\oplus {e_i})=i$. Since $n$ is odd and $2i$ is even, $n-i \neq i$. $f(X_1\oplus {e_i}) = 1$ if and only if $i=0$. Thus, we have $Sen(f,X_1)=1$ if $n=2k+1$.

If $n=4k$, $s(X_1) = 2k(4k-1)\ {\rm mod}\  4k$. Since
\begin{equation}
 \begin{aligned}
 \frac{2k(4k-1)}{4k}=&\frac{4k-1}{2}\\
 =&2k-\frac{1}{2},\\
 \end{aligned}
\end{equation}
 we have $s(X_1) = 2k$. $f(X_1\oplus {e_i})=0$ happens if and only if $s(X_1\oplus {e_i})=i$. In this case, $i=2k-i$ or $i=2k-i+4k$. Thus we have $i=k$ or $3k$.
Hence, $Sen(f,X_1)=2$ if $n=4k$.
\end{proof}

\begin{theorem}
\label{theorem3}
Let $n=p$ where $p$ is a prime number and $p>4$, then $minS(f)=1$.
\end{theorem}
\begin{proof}
It is clear that $Sen(f)=1$ at $x_0=1$ and $x_i=0, 0<i<p$. Then we have $minS(f) \leq 1$. $minS(f)=1$ implies that the equation $Sen(f,X)=0$ has no solutions.

We prove by contradiction. Suppose $Sen(f,X)=0$ has a solution $X_1$ and $j=s(X_1)$. Then we have $j\ne0$. Otherwise, $x_0$ can always flip to make $Sen(f,X_1) \geq 1$. Let $D$ be a subset in $\{0, 1, \dots, p-1\}$ such that the vector $X_1$ is viewed as the indicator function of $D$, and let $\overline{D}=\mathbb{Z}_p-D$. If $j\in D$, for each $i \in D$ and each $k \in \overline{D}$, we have $j-i\ ({\rm mod}\  p) \in D$ and $j+k \ ({\rm mod}\  p) \in D$. Similarly, if $j\in \overline{D}$, for each $i \in D$ and each $k \in \overline{D}$, we have $j-i\ ({\rm mod}\  p) \in \overline{D}$ and $j+k\ ({\rm mod}\  p) \in \overline{D}$.

Define $x\pm D=\{x\pm d, d\in D\}$. If $j\in D$, the above argument then gives that
$j-D\subseteq D$, and thus $j-D=D$. Noting that $j=s(X_1)=\sum_{d\in D} d$, sum up all the elements of the both sets $j-D$ and $D$ we obtain
$|D|j-j\equiv j\ ({\rm mod}\  p)$ and thus $|D|=2$ since $p$ is prime and $j\ne 0$. Then we have$|\overline{D}|=p-2$. For each $k \in \overline{D}$, $j+k\ ({\rm mod}\  p)$ runs over $p-2$ different values. When $p>4$, $j+k\ ({\rm mod}\  p) \in D$ does not hold for each $k \in \overline{D}$ due to $p-2>2$. Thus we deduce $j\not\in D$.

Since $j\not\in D$, $j+\overline{D} \subseteq\overline{D}$.
Thus $j+\overline{D} =\overline{D}$. Sum up all the elements of the both sets $j+\overline{D}$ and $\overline{D}$, we obtain $(p-|D|)j-j\equiv -j~({\rm mod}\ p)$ and thus $|D|=p$ since $p$ is prime and $j\ne 0$. $|D|=p$ implies $j \in D$. This is a contradiction to $j\not\in D$.

By Lemma \ref{le1} we also derive that $Sen(f,X)=0$ does not hold when $|D|=p$. Note that $p>4$ is crucial in this theorem. If $p=2$, $f(X)=0$ only at $x_0=1, x_1=1$.  If $p=3$, $f(X)=0$ only at $x_0=1, x_1=1, x_2=0$ and $x_0=1, x_1=0, x_2=1$.
\end{proof}

\section{Conclusion and Open Questions}

\begin{table}
\center
\caption{The relationship between the variable number and the minimal sensitivity}
\begin{tabular} {|c|c|}
\multicolumn{2}{c}{}\\ \hline
Variable Number&minS(f)\\ \hline
1,4,5,7,8,11,13,17,19,23&1 \\ \hline
2,3,6,9,10,12,14,15,16,18,20,21,22,24,25,26&	0\\ \hline
\end{tabular}
\end{table}
In this paper, we have explored the minimal sensitivity of a newly simplified weighted sum function. In terms of this function, we wrote a computer program which examined the relationship between the variable number and the minimal sensitivity for value $0<n<27$. The results are shown in Table 1. Other properties of the minimal sensitivity may be investigated. Related open questions are the following.
\begin{itemize}
\item It remains open whether $minS(f)=0$ always holds when $n>8$, $n$ is not a prime number.
\item It is not clear whether other kinds of weighted sum functions have similar properties of the minimal sensitivity.
\end{itemize}

\end{document}